\documentclass[conference]{IEEEtran}

\usepackage[table,dvipsnames]{xcolor}
\usepackage{times}
\usepackage{epsfig}
\usepackage{amsmath}
\usepackage{amsfonts}
\usepackage{graphicx}
\usepackage{amssymb}
\usepackage{amsthm}
\usepackage{amstext}
\usepackage{latexsym}
\usepackage{color}
\usepackage{ifthen}
\usepackage{multirow}
\usepackage{verbatim}
\usepackage{array,tabularx}
\usepackage{arydshln}
\usepackage[mathscr]{euscript}
\usepackage{accents}
 \usepackage{cite}
\usepackage{hhline}
\usepackage{caption}
\usepackage{subcaption}
\usepackage{enumerate}
\usepackage{xcolor}
\usepackage{mathtools}
\usepackage{url}
\usepackage{xparse}
\usepackage{makecell}
\usepackage{varwidth}
\usepackage{soul}

\newcommand{\CDE}{\ensuremath{C_{\star E}^{\star D}}}

\newcolumntype{C}[1]{>{\centering\let\newline\\\arraybackslash\hspace{0pt}}m{#1}}

\usepackage[normalem]{ulem}

\usepackage{amsmath,pgfplots,amssymb}
\usepackage{subcaption,graphicx}

\newcommand{\be}{\begin{equation}}
\newcommand{\ee}{\end{equation}}
\newcommand{\bear}{\begin{eqnarray}}
\newcommand{\eear}{\end{eqnarray}}
\newcommand{\bears}{\begin{eqnarray*}}
\newcommand{\eears}{\end{eqnarray*}}
\newcommand{\bi}{\begin{itemize}}
\newcommand{\ei}{\end{itemize}}
\newcommand{\ben}{\begin{enumerate}}
\newcommand{\een}{\end{enumerate}}

\usepackage{tikz}
\usetikzlibrary{shapes.geometric}
\usetikzlibrary{shapes,snakes,patterns}
\usetikzlibrary{arrows,positioning,automata,calc}
\usetikzlibrary{plotmarks}

\definecolor{ogreen}{rgb}{0,0.5,0}
\definecolor{magenta}{rgb}{1.0, 0.11, 0.81}
\definecolor{mulberry}{rgb}{0.77, 0.29, 0.55}
\definecolor{xgray}{rgb}{0.5, 0.5, 0.5}
\definecolor{ao}{rgb}{0.0, 0.5, 0.0}
\definecolor{amber}{rgb}{1.0, 0.75, 0.0}
\definecolor{capri}{rgb}{0.0, 0.75, 1.0}
\definecolor{chocolate}{rgb}{0.91, 0.41, 0.17}

\newcommand{\blue}{\color{blue}}

\newcommand{\F}{\mathbb{F}}

\DeclareMathOperator{\diag}{diag}

\newtheorem{theorem}{Theorem}

\newtheorem{conjecture}{Conjecture}

\newtheorem{example}{Example}
\newtheorem{definition}{Definition}

\newcommand*{\rowstyle}[1]{
  \gdef\@rowstyle{#1}%
  \leavevmode\@rowstyle
  \ignorespaces
}

\newcolumntype{=}{
  >{\gdef\@rowstyle{}\ignorespaces}%
}
\newcolumntype{+}{
  >{\leavevmode\@rowstyle\ignorespaces}%
}

\IEEEoverridecommandlockouts

\begin{document}
\title{Robust Private Information Retrieval from Coded Systems with Byzantine and Colluding Servers}
\author{
\IEEEauthorblockN{Razane Tajeddine\IEEEauthorrefmark{1}, Oliver W.~Gnilke\IEEEauthorrefmark{1}, David Karpuk\IEEEauthorrefmark{2}, Ragnar Freij-Hollanti\IEEEauthorrefmark{3}, Camilla Hollanti\IEEEauthorrefmark{1}\IEEEauthorrefmark{3}}
        \IEEEauthorblockA{\IEEEauthorrefmark{1} Department of Mathematics and Systems Analysis, 
		Aalto University, School of Science, 
        Espoo, Finland\\
		Emails: \{razane.tajeddine, oliver.gnilke, camilla.hollanti\}@aalto.fi}
		\IEEEauthorblockA{\IEEEauthorrefmark{2} Departamento de Matem\'aticas, 
		Universidad de los Andes, 
		Bogot\'a, Colombia\\
		Email: da.karpuk@uniandes.edu.co}
		\IEEEauthorblockA{\IEEEauthorrefmark{3} Department of Electrical and Computer Engineering, 
		Technical University of Munich, 
		Munich, Germany\\
	Email: ragnar.freij@tum.fi}
        }

\maketitle 

\begin{abstract}
A private information retrieval (PIR) scheme on coded storage systems with colluding, byzantine, and non-responsive servers is presented. Furthermore, the scheme can also be used for symmetric PIR in the same setting. 

An explicit scheme using an $[n,k]$ generalized Reed-Solomon storage code is designed,  protecting against $t$-collusion and handling up to $b$ byzantine  and $r$ non-responsive servers, when $n\geq n'=
(\nu +1) k+t+2b+r-1$, for some integer $\nu \geq 1$.
This scheme achieves a PIR rate of $1-\frac{k+2b+t+r-1}{n'}$. In the case where the capacity is known, namely when $k=1$, it is asymptotically capacity achieving as the number of files grows.

\end{abstract}

 \section{Introduction}
 
Private information retrieval (PIR) is concerned with designing schemes for a user to retrieve a certain file from a storage system without revealing the identity of the file to the servers.  This problem was introduced by Chor, Goldreich, Kushilevitz and Sudan in \cite{PIR1995}, 
where the database was viewed as an $m$-bit binary string $x = [x^1\cdots x^m]
\in \{0,1\}^m$ from which the user wants to retrieve one bit
$x^i$ while keeping the index $i$ hidden from the server. In this work, we consider files $x = [x^1 \cdots x^m]$ encoded and stored on $n$ servers, and assume that the user wants to retrieve a file $x^i$ from the storage system.

One way to achieve privacy is for the user to download all the files from the system. Of course, this scheme has a very high communication cost, or equivalently a very low PIR rate. The rate of a PIR scheme in this model is measured as the ratio of the amount of the gained information over the amount of the total downloaded data, while upload costs of the requests are usually ignored. In case of a single server storing the database, the trivial solution is the only way to guarantee \emph{information-theoretic
privacy}~\cite{PIR1995}.


\noindent{\em{Related work:}} 
Initially, PIR constructions served to reduce the total download cost from a storage system with data replicated on multiple servers \cite{beimel2001information, beimel2002breaking,dvir20142, yekhanin2010private, sun2016capacitynoncol, sun2016capacity}. 

More recently, PIR schemes were constructed on coded data. The authors in \cite{shah2014one} show that downloading one extra bit is enough to achieve privacy, if the number of servers is exponential in the number of files. In \cite{chan2014private}, the authors derive bounds on the tradeoff between storage cost and download cost for linearly  coded data. The optimal upper bounds on PIR rate were derived in \cite{banawan2016capacity}. For maximum distance separable (MDS) coded data, PIR schemes were presented in \cite{razan_salim} that achieve the asymptotic optimal download cost when the servers are non-colluding. For the case of colluding servers, the authors in \cite{freij2016private} constructed a new family of PIR schemes on MDS coded data achieving a lower download cost than the ones in \cite{razan_salim}. PIR schemes on arbitrary linear storage codes were constructed in \cite{kumar2016private}. Another line of work is symmetric PIR that was studied in \cite{wang2017secure}.


In \cite{sun2016capacity}, it is shown that the asymptotic PIR capacity for replicated data, as the number of files $m\to \infty$, for a fixed number of colluding servers $t$, is $1-\frac{t}{n}$, where $n$ is the number of nodes. 
When the data is  coded using an $[n,k]$ MDS code, it was shown in \cite{banawan2016capacity} that the asymptotic capacity is $1-\frac{k}{n}$. Codes achieving this PIR rate were first presented in~\cite{razan_salim}.

The problem of constructing PIR schemes on replicated data in which some servers can be byzantine (malicious) was considered in \cite{augot2014storage, beimel2003robust, devet2012optimally}. The asymptotic capacity of PIR on replicated storage systems with $t$ colluding servers and $b$ byzantine servers was found in \cite{banawan2017capacity} to be $1-\frac{2b+t}{n}$. In \cite{wang2017secure1}, the authors investigate the problem of providing symmetric PIR from a replicated system with colluding servers and adversaries in the system.
A robust PIR scheme on coded data with colluding and byzantine servers was constructed in \cite{zhang_ge2}. PIR from unsynchronized servers was studied in \cite{7028488}, where the files are stored on multiple servers such that some servers are not updated to the latest version, an adaptive PIR scheme is constructed for the user to retrieve privately the file he/she requires. The setting of unsynchronized servers in is similar to the byzantine servers since in both cases, some servers are responding with false, or not entirely true, information, but the work in \cite{7028488} is more restrictive and uses an adaptive scheme. 

\noindent{\em Contributions:}
In this paper, we present a general construction of robust PIR schemes with byzantine servers storing data coded with arbitrary linear $[n,k,d]$ codes. 
When the storage code is a generalized Reed Solomon code with $t$ colluding servers, $r$ unresponsive servers, and $b$ byzantine servers, if $n\geq n'=(\nu+1)k+t+2b+r-1,$ for some integer $\nu\geq 1$, we achieve a PIR rate of 
\vspace{-0.5cm}
\[1-\frac{k+t+2b+r-1}{n'}\,.\]
 
 \section{Preliminaries}
 
\subsection{Basic Definitions}

We denote the field with $q$ elements by $\mathbb{F}_q$, where $q$ is a prime power, and the set $\{1,2,\ldots , n\}$ by $[n]$. 


By an $[n,k,d]$ code we refer to a code with length $n$, dimension $k$, and minimum (Hamming) distance $d$. MDS-codes are shortly referred to as  $[n,k]$ codes, the minimum distance being implied by the fact that MDS-codes achieve the Singleton bound, \emph{i.e.}, $d=n-k+1$.

A retrieval scheme is said to be \emph{$t$-private} if the set of queries sent to any $t$-tuple of servers has zero mutual information with the identity of the desired file. In other words, a set of $t$ colluding servers cannot draw any conclusions about which file the user is downloading.


\subsection{Generalized Reed-Solomon Codes}  We propose a PIR scheme for which generalized Reed-Solomon (GRS) storage codes are naturally well-suited.  We recall the basic properties of such codes here.

\begin{definition}[GRS Codes]
	Let $\alpha=[\alpha_1\cdots \alpha_n] \in \mathbb{F}^n_q$ such that $\alpha_i \neq \alpha_j$ for all $i \neq j$, and $v=[v_1\cdots v_n] \in (\mathbb{F}_q^\times)^n$. A generalized Reed-Solomon (GRS) code of dimension $k \leq n$ is 
 \[GRS_k(\alpha,v)=\{(v_if(\alpha_i))_{1\leq i\leq n} |f \in \F_q[x], \deg(f) < k\}. \]
	\end{definition}

	The canonical generator matrix for an $[n,k]$ GRS code is given by
	\begin{equation}\label{eq:canonical} G_k(\alpha,v):=\bordermatrix{
    	& & & \cr
		{\blue{z^0}}&1 & \cdots & 1\cr
		{\blue{z^1}}&\alpha_1 & \cdots & \alpha_n \cr
		{\blue{z^2}}&\alpha_1^2 & \cdots & \alpha_n^2 \cr		
	{\textcolor{blue}{\,\vdots}}	&\vdots & \ddots & \vdots \cr
	{\textcolor{blue}{z^{k-1}}} &\alpha_1^{k-1} & \cdots & \alpha_n^{k-1} \cr
	}
	\cdot \diag(v),
	 \end{equation}
	 where the rows are indexed (in blue) by the function whose evaluation they represent and $\diag(v)$ is the diagonal matrix with the values $v_i$ on the diagonal.

\subsection{Star Products}  The star product of two codes will play an integral role in our PIR scheme, essentially determining its rate.

\begin{definition}
	Consider sub-vector spaces $V, W$ of $\F_q^n$. The star (or Schur) product $$V \star W:=\langle\{ v \star w \; : \; v\in V, w\in W\} \rangle$$ is defined as the linear span of all elements $v \star w = [v_1w_1 \cdots v_nw_n]$.
\end{definition}
For the star product of GRS codes the following useful equality holds
\[GRS_k(\alpha,v)\star GRS_\ell(\alpha,w) = GRS_{\min\{k+\ell-1,n\}}(\alpha,v\star w)\]

 \section{A PIR Scheme for Colluding Byzantine Servers}
\subsection{General Scheme}
We begin by describing a general version of our scheme that retrieves one codeword from the database ($\nu =1$). This scheme is then extended to a variant where several codewords of the database may be downloaded simultaneously ($\nu \geq 1$).

When $\nu=1$, we assume for simplicity that information is arranged in an $m \times k$ matrix $X$, where every row represents one file. Let $G_C$ be a generator matrix for the storage code $C$ and $G_{C,j}$ be the $j^{th}$ column. Then server $j$ stores the vector $y_j:=X\cdot G_{C,j}$. 
As in~\cite{freij2016private}, the queries have a random part (that does not depend on the desired file index) and a deterministic part. The random parts of the queries are generated 
by sampling $m$ random vectors from a rank $t'$ {\emph retrieval code} $D$. Let $E$ be a row vector or, equivalently, a generator matrix of a one-dimensional code. We define a third code, $\CDE=C\star D + C\star E$, by its generator matrix $\left(\begin{smallmatrix}G_{C \star D}\\G_{C \star E}\end{smallmatrix}\right)$, and denote by $d_\star$ its minimum distance. Given that
\begin{enumerate}[i)]
\item the codes $C \star D$ and $C \star E$ intersect only trivially,
\item $C \star E$ has rank $k$, and
\item $d_\star-1 \geq 2b +r$,
\end{enumerate}
 we have the following theorem.
\begin{theorem} Given codes $C$, $D$, and a vector $E$ satisfying properties i), ii), and iii) above, there is a PIR scheme that downloads an entire file from one query to each of the $n$ servers, and is correct in the presence of up to $b$ byzantine and $r$ non-responsive servers. Furthermore, this scheme is $t$-private for $t < d_{D^\perp}$.\end{theorem}

Query Construction: Queries are constructed similarly to the scheme in \cite{freij2016private}. Let $G_{D,j}$ be the $j^{th}$ column of the generator matrix of $D$ and $U \in \F_q^{m \times t'}$ a random matrix. Then the queries are given as
\begin{equation}\label{query} q_j = U G_{D,j}+E(j)e_i,\end{equation}
where $i$ is the index of the desired file, $E(j)$ denotes the $j^{th}$ entry of $E$, and $e_i$ is the $i^{th}$ standard basis vector. 

The servers' responses are $r_j=q_j.y_j$ (the inner product of the query and the server contents) for non-byzantine servers, an arbitrary element in $\F_q$ for byzantine servers, and an erasure symbol for non-responsive servers.

We prove that under the assumptions made, the $i^{th}$ row $x^i$ can be retrieved from the response vector, and any $t$-set of servers gains no information on the index $i$.
\begin{proof}
Privacy: Resistance against $t$-collusion follows analogously to the proof in \cite{freij2016private}.

Correctness: We use any decoding algorithm for the code \CDE\ to recover the vector $\rho=[q_1.y_1 \cdots q_n.y_n]$ from the possibly error and erasure carrying vector we receive from the servers. The condition iii), $d_\star-1 \geq 2b+r$, guarantees decodability. The vector $\rho$ is an element of $\CDE=C \star D \oplus C \star E$ and therefore has a unique representation as \begin{equation}\label{response}\rho=(z_1 \dots z_{k'} z'_1 \dots z'_k)\left(\begin{smallmatrix}G_{C \star D}\\G_{C \star E}\end{smallmatrix}\right)\end{equation} where $k'$ is the dimension of $C\star D$. On the other hand, the form~\eqref{query} of the queries gives a decomposition of the response vector as 
\begin{align*}
\rho &\in C\star D + (E(1)e_i.y_1\dots E(n)e_i.y_n)\\
&=C\star D + (E(1)e_i.XG_{C,1}\dots E(n)e_i.XG_{C,n})\\
&=C\star D + X_i G_{C\star E},
\end{align*} where $X_i$ is the $i^{th}$ row of the data matrix. Thus, we have $X_i=[z'_1 \cdots z'_k]$, and the requested $i^{th}$ row is the last $k$ coordinates in the representation~\eqref{response} of the response.
\end{proof}
%
Clearly, the condition that $C\star D\cap C\star E=\emptyset$ implies that that $n\geq k'+k$, where $k'$ is the dimension of $C\star D$. If $n \geq k'+ \nu k$ for $\nu >1$ we can straightforwardly extend the scheme to download more than one row, by extending $E$ to be a matrix with $\nu$ rows and requiring that
\begin{enumerate}[ii${}^*)$]
\item $C \star E$ has rank $\nu k$.
\end{enumerate}
 To apply this in order to download a single file, we must assume that the information is arranged in an $m\nu\times k$ matrix $X$, so that each file is a $\nu\times k$ matrix.

We now give an explicit description of a scheme based on GRS codes for which $k'=k+t-1$.

\subsection{Explicit Schemes from GRS Codes}
 Let $C=GRS_k(\alpha,v)$ be an $[n,k]$ GRS code and assume $t$-collusion, $b$ byzantine servers and $r$ non-responsive servers. Let $\nu$ be the maximal integer such that 
 \begin{equation} \label{eq:nbigger} n \geq (\nu +1) k+t+2b+r-1. \end{equation}
 We will download $\nu$ rows of the database. 
 Let $D=GRS_t(\alpha,w)$ be an $[n,t]$ GRS code on the same evaluation set. The rows of $E$ will be evaluations of monomials of degrees $\mu k+t-1$ for $1 \leq \mu \leq \nu$ multiplied by the diagonal matrix $\diag(w)$:\vspace{-1.2em}
 \[E=\bordermatrix{
& & & \cr
{\blue{z^{k+t-1}}} & \alpha_1^{k +t-1} & \cdots & \alpha_{n}^{ k +t-1} \cr
{\blue{z^{2k+t-1}}} & \alpha_1^{2 k +t-1} & \cdots & \alpha_{n}^{2 k +t-1} \cr
{\blue{\,\vdots}} & \vdots & \ddots & \vdots \cr
{\blue{z^{\nu k+t-1}}} & \alpha_1^{\nu k +t-1} & \cdots & \alpha_{n}^{\nu k +t-1}
}\cdot\diag(w).\]

We see now that 
$C \star D= GRS_{k+t-1}(\alpha,v\star w)$ and \\

$C\star E =$
\vspace{-1em}
\begin{small}
\begin{multline*}
 \bordermatrix{
& & & \cr
{\blue{z^{k+t-1}}} & \alpha_1^{k +t-1} & \cdots & \alpha_{n}^{ k +t-1} \cr
{\blue{z^{k+t}}} & \alpha_1^{k +t} & \cdots & \alpha_{n}^{k +t} \cr
{\blue{\,\vdots}} & \vdots & \ddots & \vdots \cr
{\blue{z^{(\nu+1) k+t-2}}} & \alpha_1^{(\nu+1) k+t-2} & \cdots & \alpha_{n}^{(\nu+1) k+t-2}
} \cdot\diag(v\star w).
\end{multline*}
\end{small}
Hence the code $\CDE=GRS_{(\nu +1)k+t-1}(\alpha,v\star w)$ is again a GRS code. It is now easy to see that properties i) and ii${}^*$) are fulfilled. Furthermore the minimum distance of $\CDE$ is given as $d_\star=n-(\nu +1) k -t+2$ which by \eqref{eq:nbigger} implies $d_\star -1\geq 2b+r$. We summarize in the following theorem.

\begin{theorem}\label{GRScodeThm}
Let the database be encoded using an $[n,k]$ GRS code $C$. Assume  $t$-collusion, $b$ byzantine, and $r$ non-responsive servers, and let $n \geq n'= (\nu +1)k+t+2b+r-1$, where $\nu$ is maximal. Then we can achieve a rate of $$\frac{\nu k}{n'}=\frac{\nu k}{(\nu +1)k +t+2b+r-1}\,.$$
This is achieved by puncturing, \emph{i.e.}, only using $n'$ servers and choosing an $[n',t]$ GRS code $D$ as the query code, and generating the queries as described above.
\end{theorem}
We venture the following conjecture.
\begin{conjecture}
The PIR capacity $\mathcal{C}$ for $t$-colluding, $[n,k]$ coded storage system with $b$ byzantine and $r$ non-responsive servers satisfies
$$\lim_{m \rightarrow \infty} \mathcal{C}=1-\frac{k+t+2b+r-1}{n}$$
as the number of files $m$ increases.
\end{conjecture}

Let us conclude this section with an example.

\begin{example}
Let $n=13, k=2, t=3, b=2,$ and $r=1$. We design a scheme with rate $R=\frac{4}{13}=1-\frac{k+t+2b+r-1}{n}$. \\
We encode using a GRS code of length $13$, using the evaluation vector $\alpha=(\alpha_1,\dots, \alpha_{13})$ and $v=\bf{1}$ the all ones vector:\vspace{-1.2em}
\[
G_C=\bordermatrix{
& & & \cr
{\blue z^0} & 1 & \cdots & 1 \cr
{\blue z^1} & \alpha_1 & \cdots & \alpha_{13} 
}.
\]
Every file consists of two rows $\left(\begin{smallmatrix} x_1^{s,1} & x_2^{s,1}\\ x_1^{s,2} &  x_2^{s,2} \end{smallmatrix}\right)$ and is encoded into two codewords $\left(\begin{smallmatrix} y_1^{s,1} \cdots y_{13}^{s,1}\\ y_1^{s,2} \cdots y_{13}^{s,2} \end{smallmatrix}\right)$.
We choose $D$ to be $GRS_3(\alpha,\bf{1})$ with generator matrix \vspace{-1.2em} \[G_D=\bordermatrix{
& & & \cr
{\blue z^0} & 1 & \cdots & 1 \cr
{\blue z^1} & \alpha_1 & \cdots & \alpha_{13} \cr
{\blue z^2} & \alpha_1^2 & \cdots & \alpha_{13}^2 
}.
\]

The matrix $E$ has to be chosen such that $\CDE$ is an error correcting code that can tolerate $1$ erasure and correct up to $2$ errors, hence $d_\star-1 \geq 5$. We pick\vspace{-1.2em} \[ E=\bordermatrix{
& & & \cr
{\blue z^4} & \alpha_1^4 & \cdots & \alpha_{13}^4 \cr 
{\blue z^6} & \alpha_1^6 & \cdots & \alpha_{13}^6
}.
\]




The code $\CDE$ is generated by the matrix\vspace{-1.2em}

\[G_{\CDE}=\bordermatrix{
& & & \cr
{\blue z^0} & 1 & \cdots & 1 \cr
{\blue z^1} & \alpha_1 & \cdots & \alpha_{13} \cr
{\blue z^2} & \alpha_1^2 & \cdots & \alpha_{13}^2 \cr
{\blue z^3} & \alpha_1^3 & \cdots & \alpha_{13}^3 \cr
{\blue z^4} & \alpha_1^4 & \cdots & \alpha_{13}^4 \cr
{\blue z^5} & \alpha_1^5 & \cdots & \alpha_{13}^5 \cr
{\blue z^6} & \alpha_1^6 & \cdots & \alpha_{13}^6 \cr
{\blue z^7} & \alpha_1^7 & \cdots & \alpha_{13}^7 \cr
}
\]
and is a $[13,8,6]$ GRS code. The queries are generated by choosing $2m$ random codewords $d^{s,t}$, $1 \leq s \leq m, \; 1\leq t \leq 2$, from $D$ and adding $E$ to the rows corresponding to the requested file $i$:
$$ (q_1\dots q_n)= \begin{pmatrix}d^{1,1}\\d^{1,2}\\ \vdots \\ d^{m,1}\\d^{m,2}\end{pmatrix}+ \begin{pmatrix}0\\[-0.5em] \vdots \\ E_1 \\ E_2 \\[-0.5em] \vdots \\ 0 \end{pmatrix}$$
The response vector is a codeword in $\CDE$ with up to $2$ errors and $1$ erasure. We decode to the vector 
$$\rho=\sum_{s=1}^m d^{s,1} \star y^{s,1}+d^{s,2} \star y^{s,2}+ E_1 \star y^{i,1}+ E_2 \star y^{i,2}.$$
Since $C\star D$ and $C \star E$ intersect trivially we can separate the part pertaining to file $i$ from the vector $\rho$. The code $C \star E$ is generated by the last four rows of $G_{{\CDE}}$. Condition iii) guarantees that we can recover $x_1^{i,1}, x_2^{i,1}, x_1^{i,2}, x_2^{i,2}$.
The achieved  rate is $R = \frac{2k}{n} = \frac{4}{13}.$\\
\end{example}

\subsection{Comparison with Previous PIR Schemes}

Recently, Zhang and Ge \cite{zhang_ge2} have constructed a PIR scheme for coded data and colluding servers, which is adaptable for unresponsive and byzantine servers (but not for both simultaneously).  In this section we briefly compare the rates obtained in this paper with those of \cite{zhang_ge2} in the asymptotic regime as $m\rightarrow\infty$.  On the one hand, the current scheme is maximally efficient when $n = n' = (\nu+1)k+t+2b+r-1$.  On the other hand, the scheme of \cite{zhang_ge2} only achieves positive rates assuming certain inequalities in the basic system parameters are satisfied, namely the obvious inequalities which guarantee that the expressions below in (\ref{badrate1}) and (\ref{badrate2}) are positive.  To compare the two schemes at their best, we grant both of these assumptions.


When $b = 0$ and $r>0$, the asymptotic rate as $m\rightarrow\infty$ from \cite{zhang_ge2} can be expressed as
\begin{equation}\label{badrate1}
\bar{R} = \frac{n}{n-r}\left(
\frac{\binom{n-r}{k}+\binom{n-t}{k}-\binom{n}{k}}{\binom{n}{k}}
\right)\,.
\end{equation}
An elementary calculation shows that $\bar{R}< \frac{n-(k+t+r-1)}{n}$, the rate obtained for the scheme described in the previous sections. In the case where $b>0$ and $r=0$, the asymptotic rate obtained in \cite{zhang_ge2} is
\begin{equation}\label{badrate2}
\bar{R} = \frac{2\left(\binom{n-b}{k}-\binom{n}{k}\right)+\binom{n-t}{k}}{\binom{n}{k}}
\end{equation}
which, again by a simple argument, is less than $\frac{n-(k+t+2b-1)}{n}\,,$ the rate obtained by the proposed scheme  in this case.

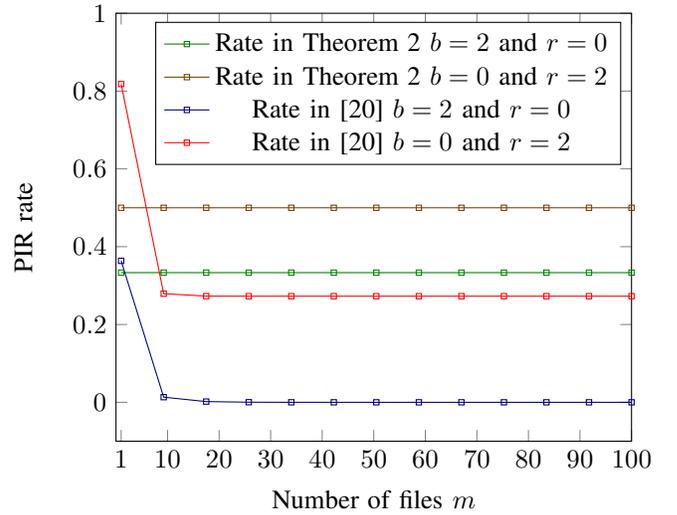
\begin{figure}
\centering
\begin{tikzpicture}

\pgfplotscreateplotcyclelist{mycolorlist}{%
blue,every mark/.append style={fill=blue!80!black},mark=*\\%
red,every mark/.append style={fill=red!80!black},mark=square\\%
brown!60!black,every mark/.append style={fill=brown!80!black},mark=otimes\\%
black,mark=star\\%
blue,every mark/.append style={fill=blue!80!black},mark=diamond\\%
red,densely dashed,every mark/.append style={solid,fill=red!80!black},mark=*\\%
brown!60!black,densely dashed,every mark/.append style={
solid,fill=brown!80!black},mark=square*\\%
black,densely dashed,every mark/.append style={solid,fill=gray},mark=otimes*\\%
blue,densely dashed,mark=star,every mark/.append style=solid\\%
red,densely dashed,every mark/.append style={solid,fill=red!80!black},mark=diamond*\\%
}
\begin{axis}[
	 xmin=0,
	 xmax=100,
     ymax=1,
    xlabel={Number of files $m$},
    ylabel={PIR rate},
    xtick={1, 10, 20,...,100},
  ]
    
   \addplot+[domain=1:100, color=green!50!black, mark = square, mark options={scale = 0.5,green!50!black},samples=13]{1-(2+3+4-1)/12};
   \addplot+[domain=1:100, color=orange!50!black, mark = square, mark options={scale = 0.5,orange!50!black},samples=13]{1-(2+3+2-1)/12}; 
   \addplot+[domain=1:100, color=blue!50!black, mark = square, mark options={scale = 0.5,blue!50!black},samples=13]{(4/11)*(1-(5/4))/(1-(5/4)^x}; 
   \addplot+[domain=1:100, color=red, mark = square, mark options={scale = 0.5,red},samples=13]{(9/11)*(1-(2/3))/(1-(2/3)^x}; 
        \legend{Rate in Theorem~\ref{GRScodeThm} $b=2$ and $r=0$, Rate in Theorem~\ref{GRScodeThm} $b=0$ and $r=2$, Rate in \cite{zhang_ge2} $b=2$ and $r=0$, Rate in \cite{zhang_ge2} $b=0$ and $r=2$}                         
  \end{axis} 
\end{tikzpicture}
\caption{\small PIR rate versus number of files $m$ when $n=12$, $k=2$, and $t=3$ following the scheme in \cite{zhang_ge2} and the scheme in this paper.}
\label{fig:comparison}
\end{figure}

Lastly, we remark that the rates obtained in \cite{zhang_ge2} decrease with an increasing number of files, while the rates we obtain are constant in the number of files.  As noted in \cite{zhang_ge2}, the rates therein outperform those of \cite{freij2016private} for a small number of files.  We can see from figure~\ref{fig:comparison} that the same holds here, but we save more precise analysis for an extended version of this paper.

\subsection{A Symmetric Variant}

A PIR scheme is \emph{symmetric} if the user, while retrieving the required file $x^i$, gains no information about any of the other files $x^{i'}$ for $i \neq i'$.  To construct a symmetric modification of our scheme, we assume the servers have access to a joint source of randomness.  This joint source of randomness outputs a uniform random codeword $s = [s_1\cdots s_n]$ of $C\star D$, and sends $s_j$ to server $j$.  

All servers then compute $r_j = q_j.y_j + s_j$, which the responsive, non-byzantine servers transmit back to the user.  As before, the user receives an erasure symbol from the non-responsive servers, and a random element of $\F_q$ from the Byzantine servers.  Also as before, the user decodes in $C_{\star E}^{\star D}$ to obtain the vector $\rho$ as in (\ref{response}), which is now easily seen to be of the form
\begin{equation}
\rho = s' + X_iG_{C\star E}
\end{equation}
where $s'$ is uniform on $C\star D$ and independent of any file indices.  Clearly $\rho$ is independent of the contents of any file $x^{i'}$ for $i'\neq i$, which guarantees symmetry.

\section{Conclusion and Future Work}

A PIR scheme was presented in this paper which can simultaneously handle coded data, colluding servers, non-responsive servers, and byzantine servers.  The scheme is an extension of previous work on PIR \cite{freij2016private} which is based on the star product of linear codes.  In the current work, the response from the servers has additional coding-theoretic properties which allow the user to correct for the erasures and errors produced by the non-responsive and byzantine servers.  The scheme has rate $1 - \frac{k+2b+t+r-1}{n'}$, where $n' = (\nu+1)k + t + 2b + r - 1$ and $\nu$ is the largest integer such that $n'\leq n$.  The scheme compares favorably to previous schemes which account for non-responsive and byzantine servers, and additionally is easily symmetrizable.

The most immediate future work will consist of generalizing this scheme so that it does not require puncturing, and achieves rate $1 - \frac{k + t + 2b + r - 1}{n}$ for any set of parameters such that $k + t + 2b + r - 1\leq n$.  Additionally, when symmetrizing our scheme, we hope to quantify how much randomness are needed for symmetrization.  We loosely conjecture that the current version is doing this with maximal efficiency, namely that $k + t -1 = \dim C\star D$ $q$-ary units of randomness are necessary.  We hope to prove this in an extended version of the current work.

\section*{Acknowledgments}
This work is supported in part by the Academy of Finland, under grants \#276031, \#282938, and \#303819 to C.~Hollanti, and by the
Technical University of Munich -- Institute for Advanced Study, funded by the German Excellence Initiative and the EU 7th Framework Programme under grant agreement \#291763, via a \emph{Hans Fischer Fellowship} held by C.~Hollanti.

O.~W.~Gnilke and R.~Tajeddine were visiting the group of Professor Antonia Wachter-Zeh at the Technical University of Munich while this work was carried out, and wish to thank for the hospitality of the LNT Chair and the COD Group.

O.~W.~Gnilke is partially supported by the Finnish Cultural Foundation.

\bibliographystyle{ieeetr}
\bibliography{coding2,coding1,references}

\end{document}